\documentclass[12pt]{article}

\usepackage{epsfig} 
 
\newcommand{\xbj}{x}

{\end{list}} 
\newcounter{enumct}

\newcommand{\captive}[1]{\rule{5mm}{0mm}%
\begin{minipage}{150mm}\caption[small]{#1}\end{minipage}}

\begin{document} 

\noindent
DESY 99-028                           \hfill ISSN 0418--9833 \\
LUNFD6/(NFFL-7167) 1999  \\

\input feynman 
\bigphotons 
\sloppy 
 
\begin{center} 
{\LARGE\bf The role of resolved virtual photons in the } \\[4mm]
{\LARGE\bf production of forward jets at HERA}
\\[4mm] 
{\Large H.~Jung, L.~J\"onsson, H.~K\"uster} \\[3mm] 
{\it Department of Physics,}\\[1mm] 
{\it Lund University, 221 00 Lund, Sweden }\\[1mm] 
{\it E-mail: jung@mail.desy.de, leif@quark.lu.se}\\[20mm] 
{\bf Abstract}\\[1mm] 
\begin{minipage}[t]{140mm} 
The measurement of forward jet cross sections has been 
suggested as a promising probe of 
new small $\xbj$ parton dynamics 
and 
the question is 
whether the new HERA data provide an indication of this. 
In this paper the influence of resolved 
photon processes has been investigated and it has been studied to 
what extent the inclusion of such processes in addition to normal 
deep inelastic scattering 
leads to agreement with data. 
It is shown that two DGLAP 
evolution chains from the hard scattering process towards the proton 
and the photon respectively, 
 are sufficient to describe effects observed 
in the HERA data, which have been attributed to BFKL dynamics. 
\end{minipage}\\[5mm] 
 
\rule{160mm}{0.4mm} 
 
\end{center} 
 
\section{Introduction} 
Experimental data from deep inelastic scattering (DIS) 
in a kinematic region where new parton dynamics is expected 
to become noticeable, i.e. at small values of the scaled proton 
momentum, $x$, are not described by models based on interactions 
with pointlike photons. 
In a previous paper~\cite{JJK_resgamma} we have demonstrated that
 the addition of 
interactions through resolved photons offers a possible explanation 
of the observed discrepancies and leads to good agreement with all 
available data. 
\par 
This paper is devoted to a more detailed discussion of the resolved 
photon concept and comparisons with data on 
forward jet production in DIS, since 
the forward jet cross section has been 
advocated as a particularly sensitive measure of 
small $\xbj$ parton dynamics~\cite{Mueller_fjets1,Mueller_fjets2}. 
Analytic calculations based on 
the BFKL equation~\cite{BFKLb,BFKLc} in the 
leading logarithmic approximation (LLA) 
 are in 
fair agreement with data. 
However, recent calculations of the BFKL kernel in 
the next-to-leading logarithmic approximation (NLLA)~\cite{BFKL_NLO} 
have given surprisingly large corrections, and it remains to be shown whether 
the data can still be reasonably described. 
\par 
Monte Carlo generators 
based on direct, point-like 
photon interactions (DIR model), 
 calculated from leading order (order $\alpha_s$) 
QCD matrix elements, and leading log parton showers 
based on the DGLAP evolution  do not 
take 
any new parton dynamics in the small $\xbj$ region 
into account and are therefore not expected 
to fit the experimental data. 
Recent results from the H1~\cite{H1_fjets_data} 
and ZEUS~\cite{ZEUS_fjets_datab} 
experiments on forward jet production 
exhibit significant deviations from the predictions of such 
models. 
Also, next-to-leading order calculations (NLO, i.e. order $\alpha_s^2$) 
assuming point-like photons predict too small a cross section compared to data. 
\par 
The study of forward jet production with contributions from 
direct as well as resolved photon processes has been performed using 
the RAPGAP~2.06~\cite{RAPGAP,RAPGAP206} Monte Carlo event 
generator. 
 
\section{Resolved Photons in DIS} 
In electron-proton scattering the internal structure of the proton as 
well as of the exchanged photon can be resolved provided the scale of 
the hard subprocess is larger than the inverse radius of the proton, 
$1/R^2_p \sim \Lambda_{QCD}^2$, and the photon, $1/R^2_{\gamma} \sim Q^2$, 
respectively. Resolved photon processes play an important role in 
photo-production of high $p_T$ jets, where $Q^2 \approx 0$, 
but they can also give considerable contributions to DIS 
processes \cite{H1_incl_jets,Chyla_res_gamma} 
if the scale $\mu^2$ of the hard 
subprocess is larger than $Q^2$, the inverse size of the photon. 
This led to the idea of including contributions from resolved photon
processes as part of deep inelastic scattering to get a more complete
description~\cite{Ingelman_edin_resgamma}. 
\par 
In the following we give a brief description of the model for resolved virtual 
photons used in the Monte Carlo generator RAPGAP. 
Given the fractional momentum transfer of the incoming electron to the 
exchanged photon, the Equivalent Photon Approximation provides the 
flux of virtual transversely polarized photons 
\cite[and references therein]{RAPGAP206,RAPGAP}. 
The contribution from longitudinally polarized photons has been neglected. 
The partonic structure of the virtual photon is defined by 
parameterizations of the parton densities, 
$x_{\gamma} f_{\gamma}(x_{\gamma},\mu^2,Q^2)$, which 
depend on the two scales $\mu^2$ and $Q^2$ 
\cite{GRS,Sasgam,Drees_Godbole}. 
The following hard subprocesses are considered (RES model): 
$ gg \rightarrow q \bar{q}$, 
$ g g \rightarrow gg$, 
$ q g \rightarrow q g $, 
$ q \bar{q} \rightarrow g g $, 
$ q \bar{q} \rightarrow q \bar{q}$, 
$ q q \rightarrow q q $. 
Parton showers on both the proton and the photon side are included. 
The generic diagram for the process $ q_{\gamma} g_{p} \rightarrow q g $ 
including parton showers is shown in Fig.~\ref{resgam1}. 
 
\begin{figure}[ht] 
\begin{center} 
\begin{picture}(30000,28000) 
\drawline\fermion[\NE\REG](5000,25000)[5000] 
\drawline\fermion[\E\REG](-2000,25000)[7000] 
\global\advance\pmidy by 500 
\put(-2000,\pmidy){$e$} 
\drawline\photon[\S\REG](5000,25000)[3] 
\global\advance\pmidx by -5000 
\put(\pmidx,\pmidy) {$y,Q^2 \to$ } 
\drawline\fermion[\E\REG](\photonbackx,\photonbacky)[1500] 
\global\advance\pmidy by 400 
\global\advance\fermionbackx by +1000 
\global\advance\pbackx by 500 
\global\advance\pbacky by -3200 
\global\advance\Yfive by + 3800 
\drawline\fermion[\S\REG](\photonbackx,\photonbacky)[1000] 
\global\advance\pmidx by -4000 
\drawline\gluon[\E\REG](\fermionbackx,\fermionbacky)[2] 
\global\Xseven = \pbackx 
\global\Yseven = \pbacky 
\global\advance\Xseven by + 500 
\global\advance\Yseven by - 300 
\put(\Xseven,\Yseven){$q_{T\;i}$} 
\global\Xone = \pbackx 
\global\Yone = \pbacky 
\global\advance\Xone by + 4500 
\global\advance\Yone by - 750 
\put(\Xone,\Yone){{\Huge 
   $\searrow$} \hspace{0.5cm} $x_{\gamma}f(x_{\gamma},\mu^2,Q^2)$ 
        \hspace{0.2cm} $\gamma$ - DGLAP} 
\drawline\fermion[\S\REG](\fermionbackx,\fermionbacky)[1500] 
 
\drawline\gluon[\E\REG](\fermionbackx,\fermionbacky)[3] 
\global\Xeight = \pbackx 
\global\Yeight = \pbacky 
\global\advance\Xeight by + 500 
\global\advance\Yeight by - 300 
\put(\Xeight,\Yeight){$q_{T\;i+1}$} 
 
\drawline\fermion[\S\REG](\fermionbackx,\fermionbacky)[2500] 
\global\advance\pmidx by -5000 
\drawline\fermion[\E\REG](\fermionbackx,\fermionbacky)[4500] 
\global\Xeight = \pbackx 
\global\Yeight = \pbacky 
\global\advance\Xeight by + 500 
\global\advance\Yeight by - 300 
\put(\Xeight,\Yeight){$p_{T}$} 
\global\advance\pmidy by 400 
\global\advance\pbackx by 500 
\global\advance\pbacky by -1500 
\global\Xsix = \pbackx 
\global\Ysix = \pbacky 
\global\advance\Ysix by + 2800 
\drawline\gluon[\S\REG](\fermionfrontx,\fermionfronty)[2] 
\global\Xtwo = \pmidx 
\global\Ytwo = \pmidy 
\global\advance\Ytwo by - 500 
\global\advance\Xtwo by - 5500 
\put(\Xtwo,\Ytwo){{$\mu^2$, $\hat{t} \to $}} 
\global\advance\Xtwo by + 6500 
 
\global\advance\Xtwo by + 6500 
\put(\Xtwo,\Ytwo){{\Huge $\} $}} 
\global\Xthree = \Xtwo 
\global\advance\Xthree by + 3000 
\global\advance\Ytwo by + 1000 
\put(\Xthree,\Ytwo){ $qg \to qg$} 
\global\Ythree = \Ytwo 
\global\advance\Ythree by - 1500 
\put(\Xthree,\Ythree){ hard scattering } 
\drawline\gluon[\E\REG](\gluonbackx,\gluonbacky)[4] 
\global\Xeight = \pbackx 
\global\Yeight = \pbacky 
\global\advance\Xeight by + 500 
\global\advance\Yeight by - 300 
\put(\Xeight,\Yeight){$p_{T}$} 
\global\advance\pmidy by 400 
\global\advance\pbacky by +400 
\global\advance\pbacky by -1500 
\drawline\gluon[\S\REG](\gluonfrontx,\gluonfronty)[3] 
\global\advance\pmidx by -3000 
\drawline\gluon[\E\REG](\gluonbackx,\gluonbacky)[3] 
\global\Xeight = \pbackx 
\global\Yeight= \pbacky 
\global\advance\Xeight by + 500 
\global\advance\Yeight by - 300 
\put(\Xeight,\Yeight){$q_{T\;j+2}$} 
\drawline\gluon[\S\REG](\gluonfrontx,\gluonfronty)[2] 
\drawline\gluon[\E\REG](\gluonbackx,\gluonbacky)[2] 
\global\Xeight = \pbackx 
\global\Yeight= \pbacky 
\global\advance\Xeight by + 500 
\global\advance\Yeight by - 300 
\put(\Xeight,\Yeight){$q_{T\;j+1}$} 
\drawline\gluon[\S\REG](\gluonfrontx,\gluonfronty)[2] 
\drawline\gluon[\E\REG](\gluonbackx,\gluonbacky)[1] 
\global\Xeight = \pbackx 
\global\Yeight= \pbacky 
\global\advance\Xeight by + 500 
\global\advance\Yeight by - 300 
\put(\Xeight,\Yeight){$q_{T\;j}$} 
\global\Xfour = \pbackx 
\global\Yfour = \pbacky 
\global\advance\Xfour by + 2500 
\global\advance\Yfour by + 2000 
\put(\Xone,\Yfour){{\Huge 
   $\nearrow$}  \hspace{0.5cm}$x_{p}f(x_{p},\mu^2)$ 
                \hspace{0.2cm} $p$ - DGLAP} 
 
\drawline\gluon[\SW\REG](\gluonfrontx,\gluonfronty)[2] 
 
\global\advance\gluonbackx by -1500 
 
\multiput(\gluonbackx,\gluonbacky)(0,-1000){3}{\line(1,0){9000}} 
\global\advance\gluonbacky by -1000 
\global\advance\gluonbackx by -500 
\put(\gluonbackx,\gluonbacky){\oval(1000,3000)} 
\global\advance\gluonbackx by +2000 
\global\advance\gluonbacky by +1000 
\global\advance\gluonbackx by -2000 
\global\advance\gluonbacky by -1000 
\global\advance\gluonbackx by -500 
\drawline\fermion[\W\REG](\gluonbackx,\gluonbacky)[2000] 
\global\advance\fermionbacky by -3000 
\global\advance\fermionbackx by 4000 
\global\advance\fermionbacky by 3000 
\global\advance\fermionbackx by -4000 
\global\advance\pmidy by 500 
\put(\pbackx,\pmidy){$p$} 
\end{picture} 
\end{center} 
\captive{Deep inelastic scattering with a resolved virtual photon and 
the $q_{\gamma} g_p \to q g $ partonic subprocess. 
\label{resgam1} } 
\end{figure} 
 
Since the photon structure function depends on the scale, $\mu^2$, of the 
hard scattering process, the cross section of resolved photon processes 
will consequently also depend on the choice of this scale. 
It has to be carefully considered in which range of $\mu^2/Q^2_0$ 
the photon-parton cross section can be factorised into a parton-parton 
cross section convoluted with the parton density of the photon. 
The parton density of the photon is evolved from a starting scale $Q^2_0$ 
 to the scale $\mu^2$, the virtuality 
at the hard subprocess, giving a resummation to all orders. 
 
\subsection{Parton Distribution Functions} 
Due to factorization of the cross section, the parton densities of both the 
virtual photon and the proton enter into the calculations. The proton 
structure function, $F_2$, has been measured to high accuracy and therefore the 
various parameterizations only give marginal differences in the 
measurable kinematic region. 
Two parameterizations of the parton distribution in the proton, 
GRV 94 HO (DIS) and CTEQ4D, have been considered, 
which both give good agreement with 
the proton structure function 
data \cite{H1_F2,ZEUS_F2}. It was found that 
the results produced were identical at the percent level, 
when keeping $\Lambda_{QCD}$ fixed. In the following we use only GRV HO. 
\par 
The photon can interact via its 
partons either in a bound vector meson state or as decoupled partons if 
the $p_T$ of the partons is high enough. 
The splitting 
$\gamma \to q \bar{q}$  is called 
the anomalous component of the photon. 
The structure function of virtual photons 
has been measured
only recently, but not nearly to the same precision as the 
proton structure function. However, it turns out that data are in good 
agreement with the parameterization of Schuler and Sj\"ostrand (SaS) 
~\cite{Sasgam}. 
The SaS parameterization 
offers a choice of $Q_0^2$ values at which the anomalous part becomes effective. 
We have studied 
these choices resulting in different magnitudes of the 
parton densities, and consequently of the cross sections. 
For the SaS 
parameterization  we have used  $Q_0^2$ as given by eq.(12) of
 ref.~\cite{Sasgam} 
 ($IP2 = 2$). This choice is also suitable for a description of other hadronic 
 final state properties (not considered in this paper), 
  like energy flow, forward particle spectra and jet 
 cross sections. 
\par 
The hadronic contribution to the virtual photon structure function 
decreases rapidly with increasing $Q^2$, which means that the main contribution
 at 
large $Q^2$ comes from the anomalous piece in the photon splitting. This 
is completely calculable in pQCD and leads to an expected agreement between 
the parameterization of Gl\"uck - Reya - Stratman~\cite{GRS} and 
that of Schuler - Sj\"ostrand~\cite{Sasgam}, but 
also with the simple ansatz of Drees - Godbole \cite{Drees_Godbole}. 
However differences 
exist in the way the hadronic part of the structure function is matched to the 
pointlike part, which just reflects the theoretical uncertainty. 
\par
In this study we will restrict ourselves to the 
SaS parton distributions~\cite{Sasgam}. 
\subsection{Choice of Scale} 
In leading order $\alpha_s$ processes,  
the renormalization scale $\mu_R$ 
and factorization scale $\mu_F$ are not well defined, 
which allows a number of reasonable choices. 
There are essentially two competing effects: 
a large scale suppresses $\alpha_s(\mu^2)$ but 
gives, on the other hand, an increased 
parton density, $xf(x,\mu^2)$, for a fixed small $x$ value. 
The net effect depends on the details of the interaction 
and on the parton density parameterization. 
\par 
In previous papers~\cite{Jung_resgamma,JJK_resgamma} 
we have tried different scales like $\mu^2= 4 \cdot p_T^2$ and $\mu^2 = Q^2 + p_T^2$, 
and found that these choices gave similar results. 
\par 
However, in resolved virtual photon processes, the choice of 
the scale $\mu^2$, at which the photon is probed, is severely 
restricted \cite{gosta_torbjorn}. 
In a partonic process $ a + b \to c + d$, where $a, b, c, d$ 
denote four-vectors and where parton $a$ has the virtuality $Q^2$, 
the transverse momentum $p_T^2$ 
of parton $c$ is given in the small angle limit ($-\hat{t} \ll \hat{s}$) by: 
$p_T^2 = \hat{s}(-\hat{t})/(\hat{s} + Q^2) $, with $\hat{s}$ and 
$\hat{t}$ being the usual 
Mandelstam variables. In a $t$ channel process the 
virtuality is given by $\mu^2 = -\hat{t}$. 
Thus we have:\footnote{We are grateful to T. Sj\"ostrand for pointing out this simple 
explanation} 
\begin{equation} 
\mu^2 = -\hat{t} = p_T^2 + Q^2 \cdot \frac{p_T^2}{\hat{s}} < Q^2 + p_T^2 
\label{kin-constraint} 
\end{equation} 
 From eq.(\ref{kin-constraint}) we see that the scale $\mu^2$ is always larger than 
the transverse momentum squared of the hard partons and less than $Q^2 + p_T^2$. 
In the following we will use $\mu^2 = Q^2 + p_T^2$ as scale for both resolved 
virtual photon processes and for direct photon processes. 
This choice of scale provides a smooth transition 
from the kinematic region of normal DIS into the range where resolved 
photons start contributing and further into the photo-production region. 
The same scale has also been used in 
NLO calculations including resolved photons in deep inelastic scattering 
\cite{Kramer_Poetter_dijets,poetter_kramer_fjets}. 
\par 
A basic test that the scale is reasonable is that the 
parton shower evolution scheme should not be able to produce partons with 
transverse momenta larger than those produced 
by the matrix element for the hard scattering process. 
 
\section{Forward Jets} 
HERA has extended the available $\xbj$ region down to values 
below $10^{-4}$, where new parton dynamics 
might show up. 
Based on calculations in the LLA of the BFKL kernel, the cross section for 
DIS events 
at low $\xbj$ and large $Q^2$ with a high $p^2_T$ jet in the 
proton direction (a forward jet) \cite{Mueller_fjets1,Mueller_fjets2} is 
expected to rise more rapidly with decreasing $\xbj$ than expected 
from DGLAP based calculations. 
New results from the H1~\cite{H1_fjets_data} and 
ZEUS~\cite{ZEUS_fjets_datab} 
experiments have recently been 
published. 
The data can be described neither by conventional DIR Monte Carlo models 
nor by a NLO calculation, while 
comparisons to analytic calculations of the 
LLA BFKL 
mechanism has proven reasonable agreement. 
\par 
It should be kept in mind that both the 
NLO calculations and the BFKL based 
calculations are performed on the parton level whereas the 
data are at the level of hadrons. 

\begin{figure}[htb] 
\begin{center} 
\epsfig{figure=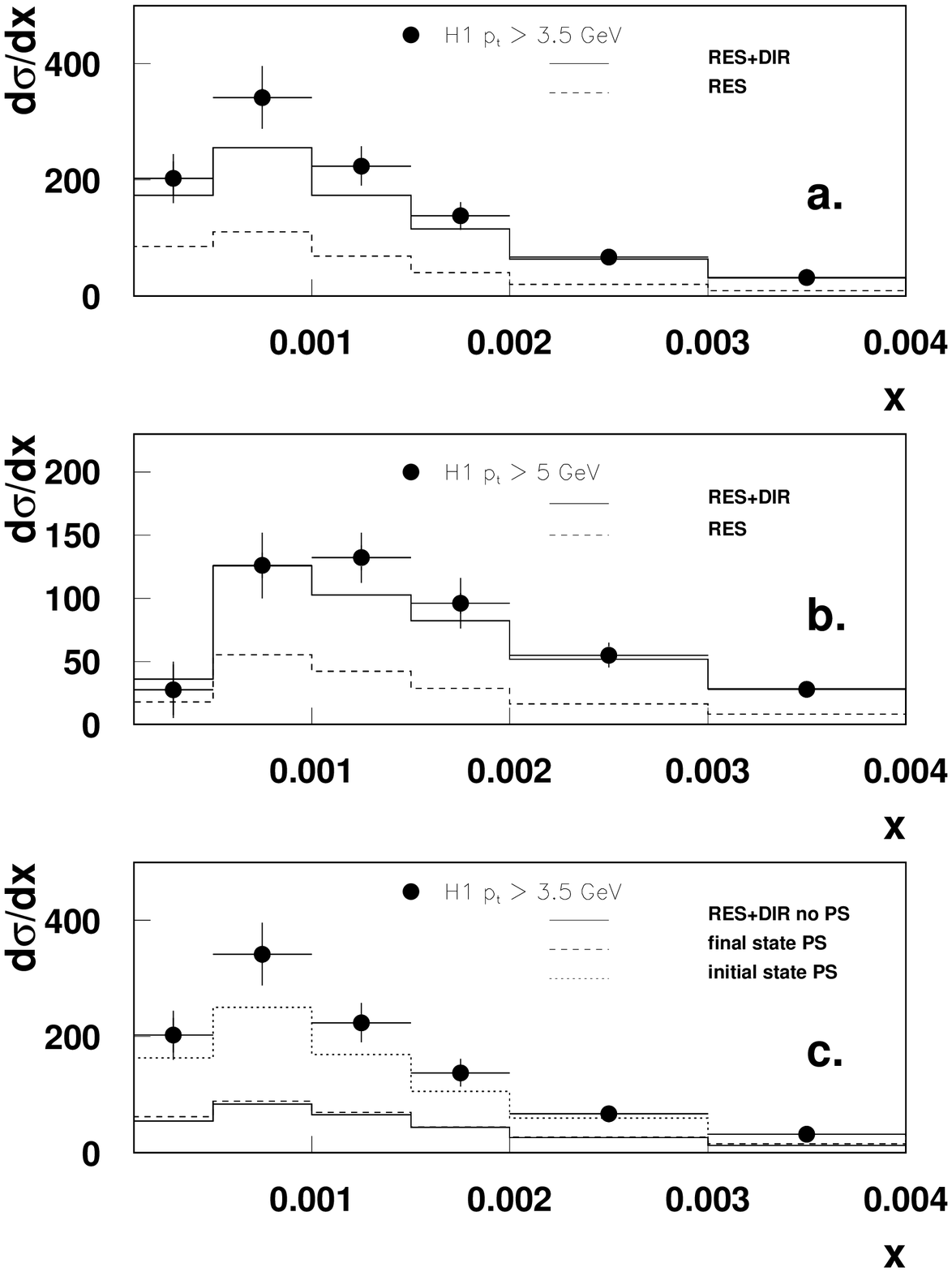, 
width=15cm,height=16cm} 
\end{center} 
\captive{ 
The forward jet cross section as a function of $\xbj$ 
for $p_{T\; jet} > 3.5$ GeV ($a$) and $p_{T\; jet} > 5$ GeV ($b$)
as measured by H1~\protect\cite{H1_fjets_data} . 
Also shown are 
the RAPGAP predictions 
for the sum 
of direct and resolved processes (solid line) as well as the 
resolved photon contribution alone (dashed line). 
The RAPGAP predictions for the sum 
of direct and resolved processes without initial and final state parton showers
(solid line), 
including only final state (dashed line) and only initial state parton showers
(dotted line) are shown in ($c$).
\label{fjet_res+dir}} 
\end{figure} 
\par 
In Fig.~\ref{fjet_res+dir}$a$ and $b$ the forward jet 
cross section as measured by the 
H1 collaboration  \cite{H1_fjets_data} is 
compared to the prediction of the RAPGAP Monte Carlo generator for both 
resolved photon process alone (labeled RES) and for 
the sum of direct and resolved processes (labeled DIR+RES). 
The calculation is performed with a 
scale $\mu^2=Q^2+p_T^2$ used in the parton densities and  
in the determination of $\alpha_s$ 
for both direct and resolved 
photon processes. Only if both processes are included, is the Monte Carlo 
able to give a good description of the measurements. 
The forward jet cross section as measured by the ZEUS experiment 
\cite{ZEUS_fjets_datab} 
can be equally well described by this Monte Carlo program using the  same 
structure functions and parameter setting. 
\par 
Recently a full NLO calculation of the forward jet cross section 
was performed \cite{poetter_kramer_fjets} with contributions 
from both direct and resolved virtual photons. Comparisons with 
H1 and ZEUS data exhibit good agreement while it is found that 
LO calculations fall below the experimental data. The assumption 
is that the parton showers in the RAPGAP model account for higher 
order effects, which would explain the good description of data 
from this model. 
\par 
As a consequence, the influence of the initial and final state QCD  
cascade on the forward jet cross section has been studied in more  
detail and compared to the calculations.  
In Fig. ~\ref{fjet_res+dir}$c$ 
we show the  
contribution to the forward jet 
cross section of H1~\cite{H1_fjets_data} coming from 
pure matrix element calculations and the effect of including 
initial and final state radiation. 
It is observed that the final state radiation gives essentially no additional 
contribution to the matrix element cross section whereas the initial state 
radiation plays an important role in bringing the Monte Carlo predictions to 
into agreement with data. 
We, thus, come to the same conclusion as in \cite{poetter_kramer_fjets}, that  
the LO order DIR+RES matrix elements are not enough to describe the measured
forward jet cross section. 
However, with initial and final state parton showers included, the data  
are described well and it seems that the parton showers are able to 
simulate contributions from higher order processes.  
We also find that the LO cross 
section calculations of \cite{poetter_kramer_fjets} agree well with  
the RAPGAP results in LO order, i.e. when parton showers are switched 
off. It should be noted that the contribution from the direct process  
with initial and final state parton showers included, corresponding to  
the difference between the RES+DIR and RES histograms in 
Fig.~\ref{fjet_res+dir}$c$, is too small to give agreement with data.  
\begin{figure}[htb] 
\begin{center} 
\epsfig{figure=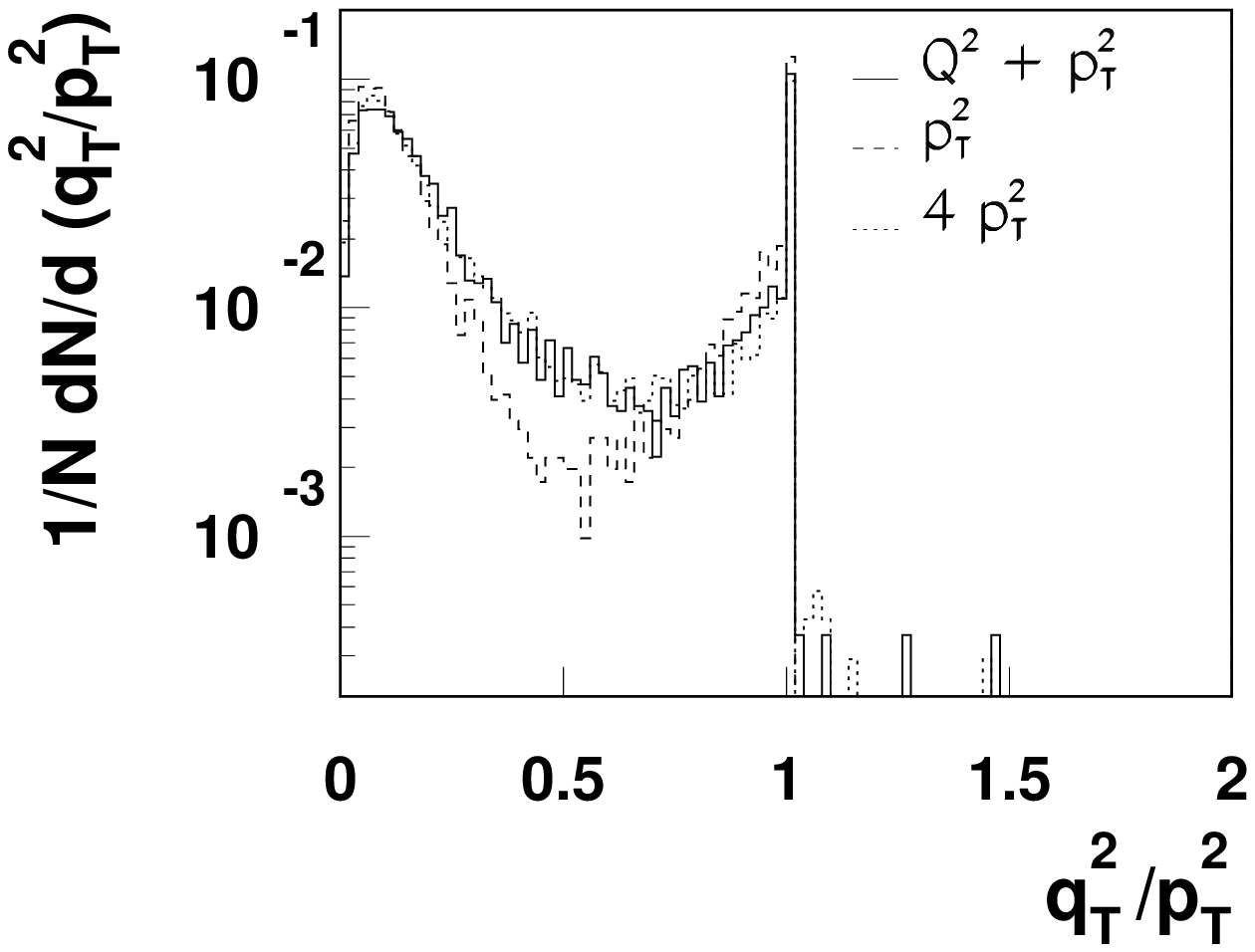, 
width=13cm,height=13cm} 
\end{center} 
\captive{The ratio $R=q^2_T/p^2_T$ 
 of the transverse momenta $q^2_T$ of partons from the initial state 
cascade to the transverse momentum $p^2_T$  
of the partons from the hard scattering 
process. The solid line corresponds to the scale $\mu^2=Q^2+p_T^2$, the dotted 
line to 
$\mu^2=4 \cdot P_T^2$ and the dashed line to $\mu^2=p_T^2$. The distribution is 
normalized to the total number of entries. Please note the logarithmic scale on 
the $y$ - axis. 
\label{fjet_ptgluon}} 
\end{figure} 
\par 
It is also important to make sure that the hardness of the scale 
doesn't lead to parton cascades which produce radiation with larger $p_T$ 
than  produced by the hard scattering matrix element. 
In Fig.~\ref{fjet_ptgluon} the ratio of the transverse momentum of any 
initial state parton $q_T^2$ to the transverse momentum $p_T^2$ 
of the hard scattering process is shown 
in the $\gamma ^* p$ CMS for events which satisfy the forward jet analysis 
criteria. The solid line corresponds to $\mu^2=Q^2 + p_T^2$, the dotted line to 
$\mu^2=4 \cdot p_T^2$ and the dashed line to $\mu^2=p_T^2$. One can see that 
essentially all partons coming from the initial state cascade have 
transverse momenta smaller than the partons of the hard scattering $p_T$, 
which is expected in a DGLAP type 
evolution, where the transverse momenta are ordered in $q_T$ towards the hard 
scattering process. Thus we conclude that the scale $\mu^2=Q^2 + p_T^2$ 
which has been used for this study fullfils the 
requirements of a DGLAP type initial state cascade. 
\begin{figure}[htb] 
\begin{center} 
\epsfig{figure=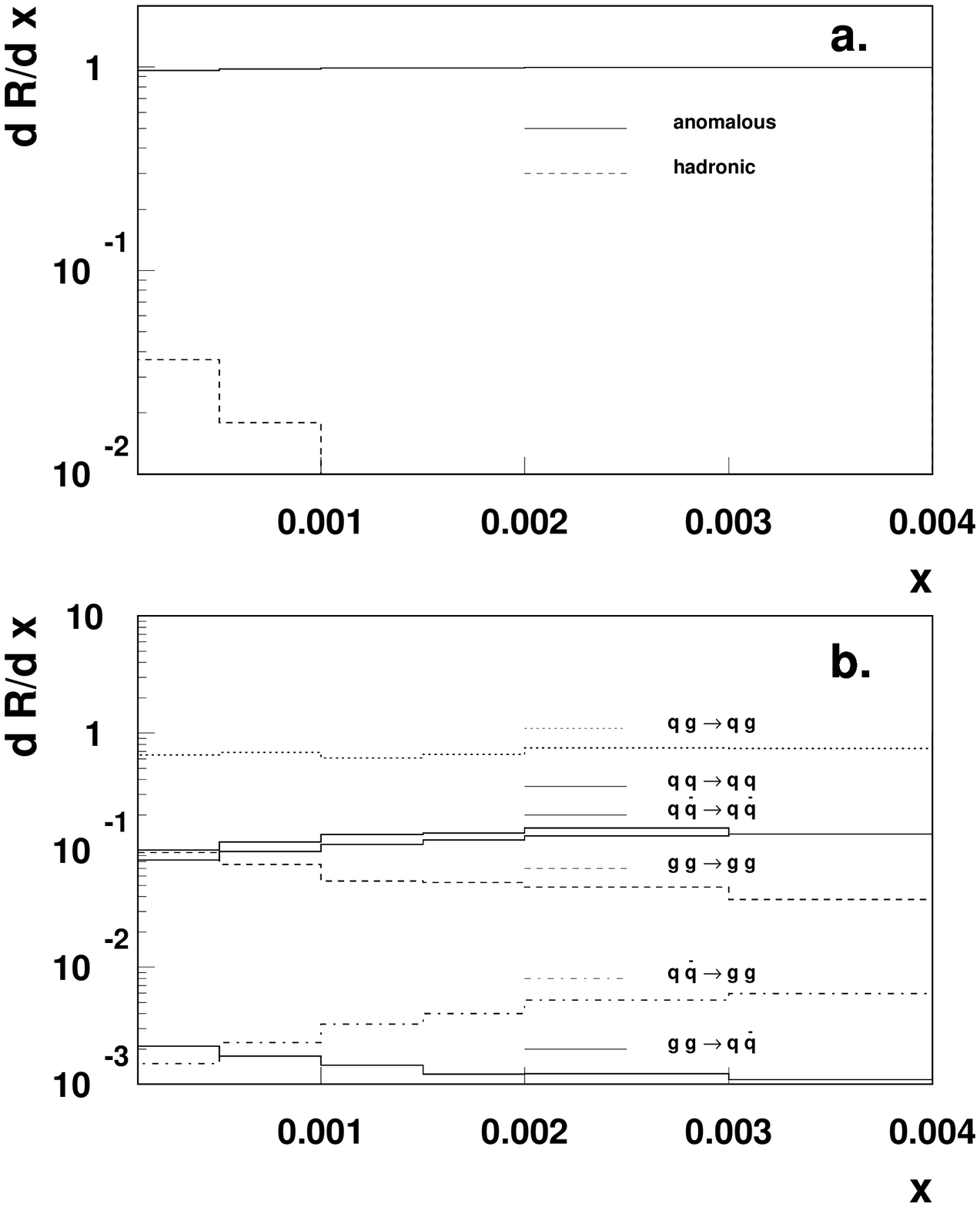, 
width=13cm,height=15cm} 
\end{center} 
\captive{Different contributions  to the total cross section of 
resolved virtual photons within the cuts of the 
forward jet analysis \protect\cite{H1_fjets_data}. 
In $(a)$ is shown 
the ratio $R= \frac{\sigma_i}{\sigma_{res.\;tot}}$, i.e.   the
 anomalous (pointlike) part (solid line) and the  
hadronic part (dashed line), respectively,
of the resolved virtual photon cross section divided by the
 the total resolved 
photon cross section as a function of $x$. 
In  
$(b)$ is shown the ratio $R=\frac{\sigma_i}{\sigma_{res.\;tot}}$ 
as a function of $x$
for different subprocesses $i$: $qg \to qg $ (dotted line),
$qq \to qq$ and $q \bar{q} \to q \bar{q}$ (upper solid line),
$gg \to gg$  (dashed line), $q \bar{q} \to gg$  (dashed-dotted line) and
$gg \to q \bar{q}$ (lower solid line).
\label{fjet_res+dir_process}} 
\end{figure}
\par 
In Fig.~\ref{fjet_res+dir_process} the different contributions 
of the total  resolved photon cross section are shown separately 
within the cuts of the forward jet analysis. 
From Fig.~\ref{fjet_res+dir_process}$a$ it is observed that the hadronic 
part of the virtual photon structure function, as expected, gives a 
negligible contribution to the measured cross section, since 
it dies off rapidly with increasing $Q^2$. 
Fig.~\ref{fjet_res+dir_process}$b$ shows that the subprocess 
$q_{\gamma}g_p \to qg$ 
contributes the most to the resolved photon cross section in the 
forward jet region ($\sim 60 \%$) 
and that  the subprocesses 
$q q  \to q q $, $q \bar{q}  \to q \bar{q} $ and $g g \to g g$ 
each give a  contribution of the order of $10 \%$. 
\par 
A small fraction of the DIS events, fulfilling the selection criteria 
for forward jets, actually contains two identified jets. 
Analytic calculations (in LLA) \cite{BFKL_dijets} 
have been performed in the same kinematic 
region and with the same jet selection as defined for the 
H1 one-jet sample. 
The predicted ratio varies from $3 \%$ to $6 \%$ as 
$\xbj$ increases from $0.5 \cdot 10^{-3}$ to $\xbj = 3 \cdot 10^{-3}$. 
Our previously reported prediction from the RAPGAP generator ~\cite{JJK_resgamma} 
including both direct and resolved photon processes was that about 1 \% 
of the total forward jet sample contains two forward jets. 
This is about a factor of 3 lower than the prediction from the 
BFKL calculations but 
a large part of this discrepancy could be due to hadronization effects 
which would reduce the prediction of the parton level BFKL calculation. 
Recently this ratio has been measured by the H1 experiment \cite{H1_fjets_data} 
to be 1 \%, in excellent agreement with the prediction of RAPGAP. 
This gives further confidence in the basic concept of resolved photons 
even at large $Q^2$. 
\par 
The ZEUS collaboration has presented a measurement of 
the forward jet cross section as a function of 
$E_T^2/Q^2$~\cite{ZEUS_fjets_pt2/q2} 
in the kinematic region $Q^2>10$ GeV$^2$, $y>0.1$, $E_{e'} > 10$ GeV, 
$\eta_{jet}<2.6$, $x_{jet} >0.036$, $E_{T\;jet}> 5$ GeV, $p_{z\;breit} > 0$ GeV, 
$2.5 \cdot 10^{-4} < x < 8 \cdot 10^{-2}$ but 
without implementing the DGLAP suppression cut 
$0.5 < E_T^2/Q^2 < 2$. The results were compared to the 
predictions from different Monte Carlo programs and the conclusion 
is that only the RAPGAP DIS generator including interactions through 
resolved virtual photons can describe the data over the full 
range in $E_T^2/Q^2$. 
\begin{figure}[htb] 
\begin{center} 
\epsfig{figure=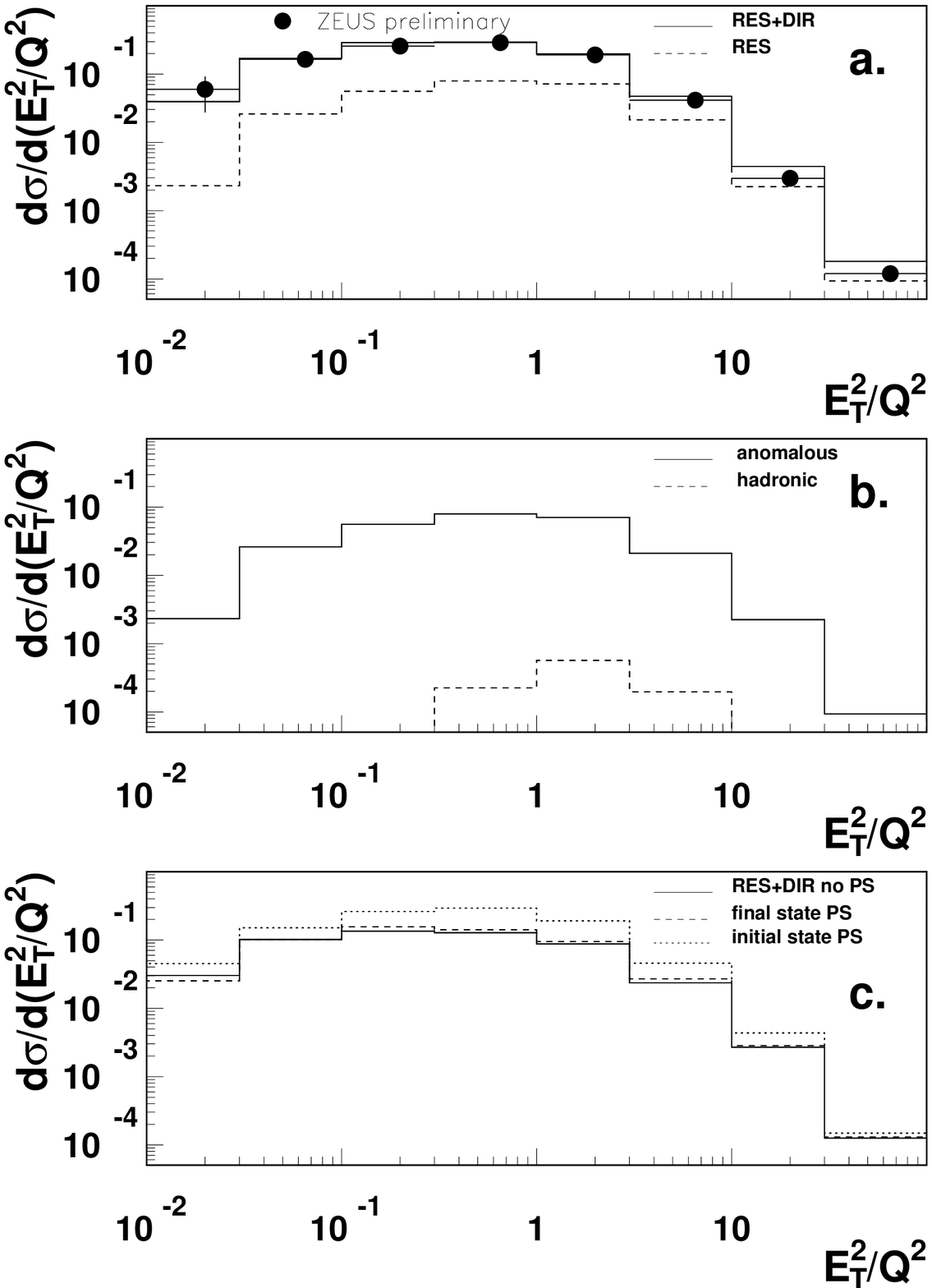, 
width=15cm,height=16cm} 
\end{center} 
\captive{The cross section of forward jets as a function of $E_T^2/Q^2$
as measured by ZEUS~\protect\cite{ZEUS_fjets_pt2/q2}. 
In $(a)$ the data are compared to  the prediction 
of RAPGAP. 
The solid line shows the sum of direct and resolved virtual photon 
contributions, 
whereas the dashed line shows the resolved  photon contribution alone. 
In $(b)$ is shown the part of 
the cross section 
coming from the anomalous component (solid line) 
and the one coming from the hadronic component of the virtual photon 
separately. 
In $(c)$ the RAPGAP predictions are shown
for the sum 
of direct and resolved processes without initial and final state parton showers
(solid line), 
including only final state (dashed line) and only initial state parton showers
(dotted line).
\label{fjet_pt2/q2}} 
\end{figure} 
In Fig.~\ref{fjet_pt2/q2}$a$ the 
ZEUS results are shown together with the prediction of RAPGAP and in 
Fig.~\ref{fjet_pt2/q2}$b$ the contributions coming from the hadronic 
and the pointlike part of the virtual photon structure function 
are presented separately. 
The influence of initial and final state radiation  
is shown in Fig.~\ref{fjet_pt2/q2}$c$ which again illustrates that the 
initial state radiation gives an essential contribution to the cross 
section while the contribution from final state radiation is negligible.  

\section{Summary and Discussion} 
 
Recent experimental data on forward jet production show deviations from 
traditional LO Monte Carlo models assuming 
directly interacting point-like photons. 
It is tempting to assume that the observed effects could be explained by 
BFKL dynamics. 
\par 
In the present study we have shown that the addition of resolved 
photon processes to the direct interactions in DIS 
leads to good agreement with the data. 
This agreement does not depend on any specific choice of scale or 
tuning of any other parameters in the RAPGAP generator. The best 
evidence of the universality of this approach is that, with the 
same parameter setting, it is possible to describe a wide range of 
other data like the transverse energy flow~\cite{H1_energyflow}, 
transverse momentum spectra of single particles~\cite{H1_ptspectra_data}, 
the (2+1) jet rate~\cite{H1_2+1jets_data} 
and single inclusive jet cross sections~\cite{H1_incl_jets}, as we have shown 
in \cite{JJK_resgamma}. 
\par 
We have observed that the dominant contributions to the resolved photon 
processes come from order $\alpha_s^2$ diagrams with the hard 
subprocess $q_{\gamma} g_p \to q g $ (see Fig.~\ref{resgam1}). 
 Since the partons which form 
the photon remnant per definition have smaller $p_T$ than the 
partons involved in the hard scattering, a situation with non $q_T$ 
ordering is created. 
\par 
In the LO DIR model, the ladder of gluon emissions is 
governed by DGLAP dynamics giving a strong ordering of 
$q_T$ for emissions between the photon and the proton vertex. 
The models describing resolved photon processes and BFKL dynamics 
are similar in the sense that both lead to a breaking of this 
ordering in $q_T$. 
The BFKL picture, however, allows for complete dis-ordering in 
$q_T$, while in the resolved photon case the DGLAP ladder is split into 
two shorter ladders, 
one from the hard subsystem to the proton vertex and one to the photon 
vertex, each of them ordered in $q_t$ (see Fig.~\ref{resgam1}). 
 Only if the ladders are long enough to produce 
additional hard radiation might it be possible to separate resolved 
photon processes from processes governed by BFKL dynamics. 
Thus the resolved photon approach may be a ``sufficiently good'' 
approximation to an exact BFKL calculation and the two approaches may prove 
indistinguishable within the range of $\xbj$ accessible at HERA. 
\par 
It should be emphasized again that the usual NLO calculation 
assuming point-like virtual 
photons contains a significant part of what is 
attributed to the resolved structure of the virtual photon in the 
LO scheme.~\cite{Kramer_Poetter_dijets}. 
The NLO calculations including contributions from direct as well as 
resolved photons, however, are similar to using LO matrix elements for 
both the photon and the proton with 
the addition of parton showers, as implemented in the RAPGAP generator. 
Recent calculations have proven good agreement with the predictions 
of RAPGAP and thus also with data. This indicates that higher order 
contributions are well simulated by the inclusion of parton showers 
in LO Monte Carlo generators.

\section{Acknowledgments} 
 
It is a pleasure to thank G. Ingelman and A. Edin for discussions 
about the concept of resolved photons. We have 
also profited from a continuous dialogue with B. Andersson, G. Gustafson 
and T. Sj\"ostrand. We want to thank G. Kramer and B. P\"otter for many 
discussions on the relation between 
 resolved photons in DIS and its relation to NLO 
calculations. 
We also want to thank M. W\"usthoff and J. Bartels 
 for discussions on BFKL and the resolved photons in DIS.

\end{document}